%% file: main.tex
\documentclass[sigconf]{acmart}

\newcommand{\E}{\mathrm{E}}
\newcommand{\Var}{\mathrm{Var}}
\newcommand{\Cov}{\mathrm{Cov}}

\usepackage{csquotes}
\usepackage{multirow}

\usepackage{enumitem}
\setlist[enumerate]{label=\arabic*.}

\begin{document}

\setcopyright{none}
\renewcommand\footnotetextcopyrightpermission[1]{}
\pagestyle{plain}

    \input{paper_header}
    \input{introduction}
    \input{background_and_related_work}
    \input{trimmed_cuped_t_test}

    \input{in_practice}
    \input{empirical_results}
    \input{conclusion}
    \bibliography{main}
    \bibliographystyle{plain}

\end{document}

%% file: paper_header.tex
\title{Improving Sensitivity in A/B Tests: Integrating CUPED with Trimmed Mean Techniques}


\author{Kevin Charette}
\orcid{0009-0001-1648-0905}
\email{kevin.charette@shopify.com}
\affiliation{%
    \institution{Shopify}
    \city{Montreal}
    \state{Quebec}
    \country{Canada}
}

\author{Tristan Boudreault}
\orcid{0009-0007-9819-7487}
\email{tristan.boudreault@shopify.com}
\affiliation{%
    \institution{Shopify}
    \city{Montreal}
    \state{Quebec}
    \country{Canada}
}

\settopmatter{printacmref=false}
\renewcommand\footnotetextcopyrightpermission[1]{}

\keywords{online controlled experiments, ab test, experiments, cuped, t-test, trimmed mean, Yuen's t-test, Welch’s t-test}

\input{abstract}

\maketitle

%% file: abstract.tex
\begin{abstract}
Accurate estimation of treatment effects in online A/B testing is challenging with zero-inflated and skewed metrics.
Traditional tests, like Welch's t-test, often lack sensitivity with heavy-tailed data due to their reliance on means, as opposed to e.g., percentiles.
The Controlled Experiments Using Pre-experiment Data (CUPED) ~\cite{cuped_original} technique improves sensitivity by reducing variance, yet
that variance reduction is insufficient for highly skewed metrics.
Alternatively, Yuen's t-test ~\cite{initial_yuen_test_article_1974} uses trimmed means to robustly handle outliers and skewness.
This paper introduces a method that combines the variance reduction of CUPED with the robustness of Yuen's t-test to
enhance hypothesis testing sensitivity.
Our novel approach integrates trimmed data in a principled manner, offering a framework that balances variance reduction
with robust location measures.
We demonstrate improved detection of significant effects with smaller sample sizes, enabling quicker experimental
decisions without sacrificing statistical power.
This work broadens the utility of controlled experiments in environments characterized by highly skewed or high-variance data.
\end{abstract}

%% file: introduction.tex
\section{Introduction} \label{sec:introduction}
A/B testing is a widely used experimental method for comparing two versions of a product or feature within organizations.
It quantifies the impact (or treatment effect) of a new version by comparing it against the existing one.
Key metrics, such as page load time or revenue per subject, are collected, and their means (or other relevant statistics) are
compared between the two versions.
These metrics frequently involve zero-inflated and right-skewed data.
A large number of subjects do not engage (zero inflation), while a few engage heavily (right skew) ~\cite{zero_inflated_right_skewed}.

This poses a challenge for the analysis of experiments using classical methods.
Commonly, Student’s t-test or Welch’s t-test are used to determine the statistical significance of the treatment effect.
These methods are however known to have low sensitivity for heavy-tailed metrics.
They become more conservative as the tail of the underlying distribution gets longer ~\cite{initial_yuen_test_article_1974}.
While the abundance of data in today's world allows organizations to recruit large cohorts for A/B testing, the cost of
running experiments long enough to achieve sufficient sample sizes can be prohibitive.
This creates an opportunity cost, limiting the number of tests conducted and delaying the rollout of optimal solutions.

Controlled Experiments Using Pre-experiment Data (CUPED) ~\cite{cuped_original} is a technique that increases the sensitivity by reducing variance.
It materially reduces the required sample size in A/B tests when pre-experiment data correlated to the metric of interest is used.
Since its introduction about a decade ago, it has been widely adopted by technology companies to run online experiments and has become a standard option for commercial platforms such as StatsSig, Eppo, Amplitude and LaunchDarkly.

However, in our experience CUPED's variance reduction is often not sufficient to reduce the sample size of heavy-tailed metrics to a practical number.
Separately from variance reduction, efforts have been put to find estimators of location that are more robust than the mean ~\cite{wilcox2010fundamentals}.
This led to trimming, another avenue to increase the sensitivity of controlled experiments when dealing with continuous variables.
This technique minimizes the influence of extreme values on the test statistics by removing a fixed percentage of the highest and lowest data points.
The trimmed mean of a population can even be preferred for its interpretation.
Its value is closer to the bulk of the distribution and represents the typical subject better than the untrimmed mean which is easily influenced by a single extreme value.
To evaluate the statistical significance of the difference in trimmed means, Yuen's t-test replaces Welch's t-test ~\cite{initial_yuen_test_article_1974}.
Truncating even a very modest fraction (0.1\%) of experiment subjects can drastically reduce the sample size needed to detect a difference.

Combining CUPED's variance reduction with trimming's robustness could yield higher sensitivity than each method on its own.
However, the CUPED literature primarily focuses on the difference in means.
No reference exists on how to leverage it with trimmed data in a rigorous way.

This paper contributes to the online experimentation and measurement science literature by introducing a new method that
combines CUPED and Yuen's t-test to increase the sensitivity of controlled experiments with heavy-tailed metrics.
We also make a contribution to the CUPED literature by showing a CUPED formulation that offers a more natural interpretation
and can be used to infer on relative treatment effects.

%% file: background_and_related_work.tex
\section{Background and related work} \label{sec:background}
Estimating the average treatment effect, denoted as $\delta$, on the population mean is the primary aim of many online
experiments.
Mathematically, the treatment effect $\delta$ for a metric $Y$ is defined as the difference between the
population means of the treatment and control groups: $\delta = \mu_2 - \mu_1$, where $\mu_j = \E[Y_j]$
represents the expected population mean of group $j$.

This section explores methods to estimate $\delta$ and methods to enhance the sensitivity and robustness of these estimates:

\begin{itemize}
    \item \textbf{Welch's t-test}: Commonly used to compare group means with unequal variances and sample sizes.
    However, its effectiveness is limited in the presence of high variance due to outliers and skewness.

    \item \textbf{Welch's t-test with CUPED}: By integrating CUPED  (Controlled Experiments Using Pre-Experiment Data)
    ~\cite{cuped_original}, we achieve variance reduction, enhancing precision.
    Yet, in certain cases, this reduction is not sufficient to attain the necessary sensitivity.

    \item \textbf{Yuen's t-test}: Utilizing trimmed means, Yuen's t-test ~\cite{initial_yuen_test_article_1974} offers a
    robust alternative by focusing on trimmed treatment effects.
    This approach mitigates the impact of outliers and skewness, providing greater robustness where traditional methods
    fall short.
\end{itemize}

This progression lays the groundwork for the forthcoming section, where we introduce a novel combination of Yuen's t-test
with CUPED.
By leveraging the strengths of both techniques, we address high-variance scenarios more effectively, achieving improved
(trimmed) treatment effect analysis.

\subsection{Welch's t-test}
Welch's t-test is the industry standard to compare the means of two groups that allows for unequal variances and sample
sizes.
More formally, it is used to test the null hypothesis that the treatment effect $\delta$ is zero versus the alternative
hypothesis that $\delta$ is different from zero.
In its most general form, Welch's t-test is defined as
\begin{align}
    t = \frac{
        \widehat{\delta}
    }{
        \sqrt{\Var\left(\widehat{\delta}\right)}
    }
    = \frac{
        \widehat{\mu}_2 - \widehat{\mu}_1
    }{
        \sqrt{\Var\left(\widehat{\mu}_2\right) + \Var\left(\widehat{\mu}_1\right)}
    }
\end{align}
where $\widehat{\delta}$ is the unbiased estimator of the treatment effect and $\Var\left(\widehat{\delta}\right)$ is the
variance of the estimator.
One of the assumptions of Welch's t-test is that both sample means should be normally distributed.
When the sample size is small, it requires that the data is normally distributed but when the sample size is large, the
central limit theorem often ensures that the numerator is normally distributed.

In its most common form, we have that $\widehat{\delta} = \overline{Y}_2 - \overline{Y}_1$ where $\overline{Y}_j$ is the
sample mean of group $j$.
This leads to the following test statistic
\begin{align}
    t = \frac{
        \overline{Y}_1 - \overline{Y}_2
    }{
    \sqrt{\Var\left(\overline{Y}_1\right) + \Var\left(\overline{Y}_2\right)}
    }, \label{eq:welch_test_statistic}
\end{align}
which follows a t-distribution with $\nu$ degrees of freedom where
\[
    \nu = \frac{
        \left( \Var\left(\overline{Y}_1\right) + \Var\left(\overline{Y}_2\right) \right)^2
    }{
        \frac{\Var^2\left(\overline{Y}_1\right)}{n_1 - 1} + \frac{\Var^2\left(\overline{Y}_2\right)}{n_2 - 1}
    }.
\]

\subsubsection{Limitations}
Because Welch's t-test relies on the sample means, its power is known to be relatively low in the presence of outliers or heavy-tailed distributions.
The presence of outliers and high skewness increases the required sample size needed to ensure that the numerator of the
test statistic be normally distributed.
When not achieved, \enquote{The result is poor probability coverage when computing a confidence interval, poor control over the
probability of a Type I error, and a biased test when using Student’s T} ~\cite{wilcox2010fundamentals}.
In practice, we had cases where one million samples were not enough to ensure that the numerator was normally distributed.
To make inferences in face of these challenges, practitioners must turn to new methods to increase statistical test power.

\subsection{Welch's t-test x CUPED} \label{subsec:cuped}
CUPED (Controlled Experiments Using Pre-Experiment Data) ~\cite{cuped_original} is a method that uses pre-experiment data
to create an unbiased treatment effect estimator $\widehat{\delta}^*$ that has lower variance than the standard estimator
$\widehat{\delta}$.
In a nutshell, CUPED explains a portion of the variance with pre-experimental data.
In this section, we cover in more detail how CUPED works which will be useful in section ~\ref{sec:trimmed_cuped_t_test}.

In the case where we have a single covariate, denoted X, the CUPED adjusted treatment effect estimator is
\begin{align*}
    \widehat{\delta}^* = \overline{Y}_2^* - \overline{Y}_1^*,
\end{align*}
where
\begin{align}
    \overline{Y}_j^* = \overline{Y}_j - \theta \overline{X}_j \label{eq:cuped_mean_residual}
\end{align}
is the CUPED mean residual of group $j$.
Through the paper, we use the $Y^*$ notation to represent the CUPED related variables.
Assuming that the pre-experimental distribution of the covariate is the same for both groups, we have that the CUPED
adjusted treatment effect estimator $\widehat{\delta}^*$ is unbiased for $\delta$.

The variance of the CUPED mean residual is given by
\begin{align}
    \Var\left(\overline{Y}_j^*\right) = \Var\left(\overline{Y}_j\right)
        + \theta^2 \Var\left(\overline{X}_j\right)
        - 2\theta \Cov\left(\overline{Y}_j, \overline{X}_j\right) \label{eq:variance_cuped_mean_residual}
\end{align}
and the variance of the CUPED adjusted treatment effect estimator is given by
\begin{align*}
    \Var\left(\widehat{\delta}^*\right) &= \Var\left(\overline{Y}_2^*\right) + \Var\left(\overline{Y}_1^*\right)
\end{align*}
because the two groups are independent.
It can be shown ~\cite{deng2023augmentationdecompositionnewlookcuped} that the value of $\theta$ that minimizes the variance of $\widehat{\delta}^*$ is given by
\begin{align}
    \theta = \frac{
        \Cov\left(\overline{X}_2, \overline{Y}_2\right) + \Cov\left(\overline{X}_1, \overline{Y}_1\right)
    }{
        \Var\left(\overline{X}_2\right) + \Var\left(\overline{X}_1\right)
    }. \label{eq:cuped_theta}
\end{align}
The CUPED treatment effect estimator is unbiased and has a smaller variance than the original estimator.
\cite{cuped_original, deng2023augmentationdecompositionnewlookcuped} have shown that the variance reduction improves
with the correlation between the covariate and the metric of interest.

In order to use Welch's t-test with CUPED, one would compute the CUPED mean residuals and their associated variances and
then use Welch's t-test on the CUPED mean residuals.

\subsubsection{Limitations of CUPED}
In practice, many experiments and metric combinations lack pre-experiment data or have pre-experiment data that is
uncorrelated with the metric of interest.
When this occurs, the benefits of CUPED are lost, and we are back to the unadjusted Welch's t-test performance.

Also, there are some cases when the distribution of each group is so skewed and the variance so high that even with a
massive variance reduction, we would still need a very large sample size to achieve enough sensitivity or to ensure that
the numerator of the test statistic is normally distributed.
When this happens, it is time to look at using more robust estimators of location that are less sensitive to the presence
of outliers and skewness.

\subsubsection{CUPED as an estimator adjustment}
We are presenting CUPED as an adjustment of the treatment effect estimator.
However, many blogs and papers citing CUPED interpret it as a regression adjustment method, defining it in
terms of averages of individual residuals ~\cite{deng2023augmentationdecompositionnewlookcuped}.
We deliberately choose to work at the treatment effect level since the CUPED authors developed the theory directly on estimators, instead of modeling subject-level data points like regression models do.
Furthermore, interpreting CUPED as a regression adjustment method makes it challenging to combine it
with Yuen's t-test as presented later in this paper.

\subsection{Trimmed mean} \label{subsec:trimmed_mean}
The trimmed mean has been used and studied in the field of statistics for a long time as a robust alternative to the
mean ~\cite{1968DixonTukeyTrimming2, 1963TukeyMcLaughlinTrimming, initial_yuen_test_article_1974}.
The idea is to remove a certain percentage of the data from both ends of the distribution and then compute the mean of
the remaining data.

Literature on the trimmed mean often give two keys reasons to favor it over conventional means:
enhanced statistical power in hypothesis testing and location closer to the mode of skewed distributions ~\cite{wilcox1994tuckey_mcLaughlin_yuen_study}.

Let $X$ be any random variable with density $f(x)$, the trimmed distribution is defined as
\[
    f_t(x) = \frac{f(x)}{1 - 2\gamma}, \qquad {x_\gamma \leq x \leq x_{1 - \gamma}},
\]
where $0 < \gamma < .5$ represents the desired amount of trimming on each end and where $x_\gamma$ and $x_{1 - \gamma}$ are the
$\gamma$ and $1 - \gamma$ quantiles of the non-trimmed distribution.
The trimmed  mean $\mu_t$ is the mean of the distribution after it has been trimmed.

A natural and unbiased estimator for $\mu_t$ is the sample trimmed mean $\overline{Y}_t$ ~\cite{wilcox2012introduction}.
Let $Y_1, Y_2, \ldots, Y_n$ be a random sample from a distribution and $Y_{(1)}, Y_{(2)}, \dots, Y_{(n)}$ the ordered
sample. The sample trimmed mean is defined as
\begin{align}
    \widehat{\mu}_{t} =  \overline{Y}_t =  \frac{1}{n-2g} \sum_{i = g + 1}^{n - g} Y_{(i)}, \label{eq:sample_trimmed_mean}
\end{align}
where $g$ is the number of observations to be trimmed from each end of the distribution defined as
$g = \left[ \gamma n \right]$, the greatest integer less than or equal to $\gamma n$.

In order to make some useful inference on $\mu_t$, we need to estimate the variance of the trimmed mean estimator
$\widehat{\mu}_t$.
The problem with the sample trimmed mean is that it is a linear combination of dependent random variables,
namely a linear combination of the order statistics ~\cite{wilcox2012introduction}.
This complexifies how we estimate the variance of the sample trimmed mean compared to the sample mean.
Previous work ~\cite{wilcox2012introduction, initial_yuen_test_article_1974, staudte2011robust} has shown that the variance of the trimmed mean can be estimated with
\begin{align}
    \widehat{\Var\left(\overline{Y}_t \right)} =
        \frac{\left(n-1\right)}{\left( h-1 \right)} \frac{s^2_w}{h}, \label{eq:standard_error_trimmed_mean}
\end{align}
where $h = n - 2g$ is the effective sample size and where $s^2_w$ is the winsorized sample variance defined as
\begin{align}
    s^2_w = \frac{1}{n-1} \sum_{i = 1}^{n} \left(W_i - \overline{W}\right)^2, \label{eq:winsorized_sample_variance}
\end{align}
and where $W_i$ is the winsorized sample defined as
\begin{align}
    W_i =
    \begin{cases}
        Y_{(g + 1)}, & \text{if}\ Y_i \leq Y_{(g + 1)} \\
        Y_i, & \text{if}\ Y_{(g + 1)} < Y_i < Y_{(n - g)} \\
        Y_{(n - g)}, & \text{if}\ Y_i \geq Y_{(n - g)}
    \end{cases} \label{eq:winsorized_sample}
\end{align}
and finally where
\begin{align}
    \overline{W} = \frac{1}{n} \sum_{i = 1}^{n} W_i. \label{eq:winsorized_sample_mean}
\end{align}
Similarly, we can estimate the covariance between 2 trimmed means with
\begin{align}
    \widehat{\Cov\left(\overline{Y}_{t1}, \overline{Y}_{t2}\right)} & = \frac{\hat{\sigma}_{12w}}{h} \nonumber \\
    & = \frac{
        \sum_{i=1}^n\left(W_{i1} - \overline{W}_1\right) \left( W_{i2} - \overline{W}_2\right)
    }{h(h-1)} \label{eq:standard_error_trimmed_mean_cov}
\end{align}
Finally, it was shown ~\cite{stigler_1973_asymptotic_trimmed_mean} that the sample trimmed mean is asymptotically normal.

If we summarize, the sample trimmed mean is a robust estimator of the population trimmed mean and is asymptotically normal.
These properties are similar to the necessary ingredients of Welch's t-test.
However, to compare the trimmed means of two groups, a new statistical test is required.

\subsection{Yuen's t-test: Welch's t-test $\times$ Trimmed mean} \label{subsec:yuen_test}
Yuen's t-test ~\cite{initial_yuen_test_article_1974} is a robust version of the two samples Welch's t-test that is based on
the sample trimmed mean.
It tests the hypothesis that, for a given $\gamma$, the trimmed treatment effect $\delta_t = \mu_{t2} - \mu_{t1}$ is 0 versus the alternative hypothesis that
$\delta_t$ is different from 0.
In Yuen's t-test, each group is trimmed separately.
Similar to Welch's test statistic ~\eqref{eq:welch_test_statistic}, the test statistic is defined as
\begin{align}
    t &= \frac{
        \overline{Y}_{t2} - \overline{Y}_{t1}
    }{
        \sqrt{\widehat{\Var\left(\overline{Y}_{t2}\right)} + \widehat{\Var\left(\overline{Y}_{t1}\right)}}
    }, \nonumber \\
      &= \frac{\overline{Y}_{t2} - \overline{Y}_{t1}}{\sqrt{d_1 + d_2}}, \label{eq:yuen_test_statistic}
\end{align}
where
\begin{align*}
    d_j &= \widehat{\Var\left(\overline{Y}_{tj} \right)} = \frac{\left( n_j - 1 \right)s^2_{wj}}{(h_j - 1) h_j}.
\end{align*}
The null distribution of the test statistic is a t-distribution with
\begin{align}
    \hat{\nu} = \frac{\left( d_1 + d_2 \right)^2}{\frac{d_1^2}{h_1 - 1} + \frac{d_2^2}{h_2 - 1}} \label{eq:degrees_of_freedom_yuen}
\end{align}
degrees of freedom.

\subsubsection{Benefits}
Yuen's t-test offers several advantages over Welch's t-test.
In the case of heavy-tailed data, Yuen's t-test power is higher than Welch's t-test and its false positive rate is also
closer to the nominal rate ~\cite{initial_yuen_test_article_1974}.
Furthermore, by focusing on trimmed treatment effects it provides insights into how more typical subjects are impacted by a treatment.
The trimmed treatment effect estimator is also more robust to outliers than the untrimmed estimator.

\subsubsection{Limitations}
Yuen's t-test estimates the treatment effect on a narrower segment of the population ($\delta_t$) rather than the
overall population ($\delta$), which may yield different results when distributions are skewed ($\delta \neq \delta_t$).
In instances where a treatment affects the distribution tails differently from the main body, there is a risk of making
suboptimal decisions, such as implementing a treatment that may not benefit the most valuable subjects.

Nonetheless, in most cases, when the tails are included, test sensitivity is too low, leading to a trade-off between
learning about the majority of the population or learning nothing at all.

It is worth noting that Yuen's t-test is not always more powerful than Welch's t-test.
For skewed distributions, methods based on means might show more power due to potentially larger differences between
population means compared to trimmed means.
If the trimmed effect is smaller than the non-trimmed effect and the variance reduction from trimming insufficient,
Welch's t-test may be more powerful.
However, practical experience suggests this situation is very rare.
On the contrary, practitioners often encounter cases where traditional mean comparison fails to detect a significant difference between groups, while trimmed mean comparison successfully finds one ~\cite{wilcox2010fundamentals}.

%% file: trimmed_cuped_t_test.tex
\section{Yuen's t-test with CUPED} \label{sec:trimmed_cuped_t_test}
In order to combine CUPED and Yuen's t-test, we need to find a CUPED adjusted estimator
$\widehat{\delta}^*_{t} = \overline{Y}^*_{t2} - \overline{Y}^*_{t1}$ that is unbiased for the true trimmed effect
$\delta_t = \mu_{t2} - \mu_{t1}$ with smaller variance than the original estimator $\widehat{\delta_t} = \overline{Y}_{t2} - \overline{Y}_{t1}$.

Once we found that estimator, we can use Yuen's t-test to evaluate the null hypothesis that the trimmed treatment effect is
zero by using
\begin{align}
    t^* = \frac{
        \overline{Y}^*_{t2} - \overline{Y}^*_{t1}
    }{
        \sqrt{\Var\left(\overline{Y}^*_{t2}\right) + \Var\left(\overline{Y}^*_{t1}\right)}
    } \label{eq:cuped_adjusted_t_test_statistic},
\end{align}
where this new $t^*$ statistic is more sensitive than the original $t$ statistic based on the trimmed mean.

\subsection{Trimming CUPED adjusted data is not possible}
The most natural way of combining CUPED and trimming would have been to run Yuen's t-test on the CUPED subject-level
adjusted data.
In other words, the CUPED adjustment is applied to individual subjects, followed by trimming a fraction of the top and
bottom subjects based on the adjusted metric.
However, this approach results in biased estimates of the trimmed treatment effect.

To get an intuition why, let's consider a familiar and related measure of location: the median.
Under general conditions, the difference of marginal population medians is not equivalent to the median of paired differences.
This same non-equivalence property applies to trimmed means ~\cite{wilcox2012introduction}.
That is, if $Y^{*}_{ij} = Y_{ij} - \theta X_{ij}$ is the CUPED residual for subject $i$ of group $j$ and
$Y^{*}_{tij}$ is the trimmed CUPED residual, then under general conditions
\begin{align}
    \E[Y^{*}_{tij}] \neq \E[Y_{tij} - \theta X_{tij}] \label{eq:individual_trimmed_cuped_bias}.
\end{align}
In order to remain unbiased, the trimmed CUPED estimator $\widehat{\delta}_{t}^{*}$ requires that
$\E[Y^{*}_{ti2} - Y^{*}_{ti1}] = \delta_t$.
However, if we were to trim the CUPED adjusted data $Y^{*}_{ij}$ we would have that
\begin{align*}
    \E[Y^{*}_{ti2} - Y^{*}_{ti1}]  & \neq \E\left[Y_{ti2} - \theta X_{ti2}\right] - \E\left[Y_{ti1} - \theta X_{ti1}\right] \\
    & = \delta_t,
\end{align*}
where the first inequality results from equation ~\eqref{eq:individual_trimmed_cuped_bias}.
Therefore, trimming the individually adjusted CUPED data would result in a biased estimator.

\subsection{CUPED adjusted trimmed treatment effect estimator $\widehat{\delta}^*_{t}$} \label{subsec:trimmed_cuped_treatment_effect_estimator}
In order to ensure that the CUPED adjusted trimmed treatment effect estimator is unbiased, we first independently trim
the means for the variable of interest and the covariate and then we use CUPED to reduce the variance.
This leads to the following CUPED adjusted trimmed effect estimator:
\begin{align*}
    \widehat{\delta}^*_{t} = \overline{Y}_{t2}^* - \overline{Y}_{t1}^*,
\end{align*}
where
\begin{align*}
    \overline{Y}_{tj}^* = \overline{Y}_{tj} - \theta \overline{X}_{tj}
\end{align*}
is the CUPED trimmed mean residual for group $j$.
Recalling that, for a given $\gamma$, $\E\left[\overline{X}_{t1}\right] = \E\left[\overline{X}_{t2}\right]$ because the covariate distributions
are identical before the start of the experiment, it is easy to see that $\widehat{\delta}^*_{t}$ is unbiased for $\delta_{t}$:
\begin{align*}
    \E\left[\widehat{\delta}^*_{t}\right] & = \E\left[\overline{Y}_{t2}^*\right] - \E\left[\overline{Y}_{t1}^*\right] \\
    & = \mu_{t2} - \theta \mu_{tx2} - \mu_{t1} + \theta \mu_{tx1} \\
    & = \mu_{t2} - \mu_{t1} \\
    & = \delta_{t}.
\end{align*}
From there, everything that was presented in section ~\ref{subsec:cuped} and in the original CUPED paper ~\cite{cuped_original}
can be reused by replacing the CUPED mean residual $\overline{Y}_{j}^*$ with the CUPED trimmed mean residual
$\overline{Y}_{tj}^*$.
In particular, we have that the variance of the CUPED trimmed mean residual for group $j$ is
\begin{align}
    \Var\left( \overline{Y}^{*}_{tj} \right) & = \Var\left( \overline{Y}_{tj} - \theta \overline{X}_{tj} \right) \nonumber \\
    & = \Var\left( \overline{Y}_{tj} \right) +
    \theta^2 \Var\left( \overline{X}_{tj} \right) -
    2 \theta \Cov\left( \overline{Y}_{tj}, \overline{X}_{tj} \right) \label{eq:variance_cuped_adjusted_trimmed_mean_residual}
\end{align}
and that the variance of the CUPED adjusted trimmed effect estimator is
\begin{align}
    \Var\left( \widehat{\delta}^*_{t} \right) & = \Var\left( \overline{Y}_{t2}^* \right) + \Var\left( \overline{Y}_{t1}^* \right), \nonumber
\end{align}
which is minimized when
\begin{align}
    \theta = \frac{
        \Cov\left(\overline{X}_{t2}, \overline{Y}_{t2}\right) + \Cov\left(\overline{X}_{t1}, \overline{Y}_{t1}\right)
    }{
        \Var\left(\overline{X}_{t2}\right) + \Var\left(\overline{X}_{t1}\right)
    }. \label{eq:trimmed_cuped_theta}
\end{align}

\subsubsection{Losing Subject-Level Data}
Trimming here is done for $Y$ (outcome of interest) separately from $X$ (pre-experimental outcome).
One of the consequences of trimming the variables in isolation is that we must abandon subject level data and work in
aggregates.
After trimming, the individual level data is sparse.
For some subjects the pre-experiment observation $X_{ij}$ will be cut by trimming, but not the in-experiment observation
$Y_{ij}$ or vice versa.
Only keeping complete data for both $X$ and $Y$ is not an option as it would introduce biases in the trimmed mean estimate.
This is because the remaining subjects would not be representative of the trimmed population for the covariate or the
variable of interest.

One impact of this is how we compute the standard error of the CUPED adjusted trimmed mean residual
$\overline{Y}^{*}_{tj}$.
When using CUPED on subject-level non-trimmed data, we typically compute the individually adjusted $Y_{ij}^{*}$ and then
compute the sample variance of the population from which we find the standard error.
However, this approach fails as we no longer have complete individual-level data.
In order to estimate the standard error, we need to deconstruct the variance of the trimmed CUPED mean residual and
compute it using ~\eqref{eq:variance_cuped_adjusted_trimmed_mean_residual}.

\subsection{Step-by-step implementation}
The goal of this section is to explain how to combine the Yuen's t-test and CUPED once you gathered the sample of data.
Looking at the CUPED adjusted Yuen's t-test $t^*$ statistic ~\eqref{eq:cuped_adjusted_t_test_statistic}, we need to
\begin{enumerate}
    \item Estimate the sample trimmed mean $\overline{X}_{tj}$ and $\overline{Y}_{tj}$ using ~\eqref{eq:sample_trimmed_mean} for both groups.
    \item Estimate the trimmed mean variance $\widehat{\Var\left( \overline{Y}_{tj}\right)}$ and $\widehat{\Var\left( \overline{X}_{tj}\right)}$ using ~\eqref{eq:standard_error_trimmed_mean} for both groups.
    \item Estimate the trimmed mean covariance $\widehat{\Cov\left( \overline{X}_{tj}, \overline{Y}_{tj}\right)}$ using ~\eqref{eq:standard_error_trimmed_mean_cov} for both groups.
    \item Estimate $\theta$ using ~\eqref{eq:trimmed_cuped_theta} and by replacing the variance and covariance by their
        estimations computed above.
    \item Compute the CUPED trimmed mean residuals $\overline{Y}_{tj}^*$ for both groups.
    \item Estimate the standard error of the CUPED trimmed mean residual $\widehat{\Var\left( \overline{Y}^{*}_{tj} \right)}$ using ~\eqref{eq:variance_cuped_adjusted_trimmed_mean_residual} for both groups.
    \item Compute Yuen's CUPED adjusted $t^*$ statistics using ~\eqref{eq:cuped_adjusted_t_test_statistic}.
    \item Compute the degrees of freedom using ~\eqref{eq:degrees_of_freedom_yuen}.
    \item Perform Yuen's t-test using the $t^*$ statistic and the degrees of freedom.
\end{enumerate}

%% file: in_practice.tex
\section{Enhanced CUPED Estimator}
When applying CUPED, whether for trimmed means or regular sample means, we favor a slightly modified adjustment to the
one presented in the original CUPED article~\cite{cuped_original} and defined in equation
~\eqref{eq:cuped_mean_residual}. The preferred adjustment is defined as:
\begin{align*}
    \overline{Y}^{**}_{j} = \overline{Y}_{j} - \theta \left( \overline{X}_{j} - \overline{X} \right)
\end{align*}
where $\overline{Y}_{j}$ and $\overline{X}_{j}$ are the sample mean for group $j$ and $\overline{X}$
is the pooled sample mean for the covariate.
In this setup, we have that $\E[\overline{Y}^{**}_{j}] = \mu_{j}$ because
$\E[\overline{X}_j]$ = $\E[\overline{X}]$.
This contrasts with the original definition of $\overline{Y}^{*}_{j}$ defined at
~\eqref{eq:cuped_mean_residual} where $\E[\overline{Y}^{*}_{j}] = \mu_{j} - \theta \mu_{x}$.
This estimator is more convenient because $\overline{Y}^{**}_j$ can now be seen as the CUPED adjusted mean for group $j$.

In the trimmed mean context, we can use a similar change to get the CUPED adjusted trimmed mean:
\begin{align}
    \overline{Y}^{**}_{tj} = \overline{Y}_{tj} - \theta \left( \overline{X}_{tj} - \overline{X}_{t} \right) \label{eq:improved_cuped_adjusted_trimmed_mean}
\end{align}
where $\overline{Y}_{tj}$ and $\overline{X}_{tj}$ are the trimmed mean for group $j$ and $\overline{X}_{t}$
is the pooled trimmed mean for the covariate.
Similarly to the untrimmed case, we have that $\E[\overline{Y}^{**}_{tj}] = \mu_{tj}$.

\subsection{Benefit 1: Natural interpretation}
When reporting the results, it is now possible to report and interpret the CUPED adjusted mean $\overline{Y}^{**}_{j}$
instead of or in addition to the sample mean $\overline{Y}_{j}$.
As originally defined, $\overline{Y}^{*}_{j}$ did not have much reporting value as it represented the mean of the residuals
when regressing $Y$ on $X$ and was difficult to interpret.

\subsection{Benefit 2: Enables inference on relative change}
Making inference on the relative change of the mean is very common in A/B testing.
Instead of making inferences on $\delta = \mu_2 - \mu_1$, people often prefer to make inferences on the relative change
defined as $\delta_{\%} = \frac{\mu_2 - \mu_1}{\mu_1}$.
Many methods ~\cite{Fieller1940, Fieller1954, RonnyKohaviPercentChange2009, DengPercentChangeDelta} have been proposed
to make inferences on the percentage change of the mean.
All of these methods require unbiased estimators of $\mu_2$ and $\mu_1$ and not just an unbiased
estimator of $\delta$ which is what the initial CUPED method provides.

By using the CUPED adjusted mean $\overline{Y}^{**}_{j}$ proposed in this section, we now have access to an unbiased
estimator of $\mu_{j}$ with reduced variance.
This allows us to make inference on the relative change of the mean using the same methods as the one used for the mean.

\section{Yuen’s t-test and CUPED in Practice}
\subsection{Choosing the Trim Percentage $\gamma$}
The optimal percentage of data to trim is its own topic.
If the data is skewed and $\gamma > 0$, we typically observe the following
\begin{itemize}
    \item $\mu_{j} \neq \mu_{tj}$: the population mean is not equal to the trimmed population mean.
    \item $\E[\widehat{\mu}_{tj}] = \mu_{tj}$: the trimmed sample mean estimator is unbiased for the trimmed population mean.
    \item $\Var[\widehat{\mu}_{tj}] < \Var[\widehat{\mu}_{j}]$: the variance of the trimmed sample mean estimator is smaller than the variance of the sample mean estimator.
    \item $\left| \delta - \delta_t \right| = \Delta > 0 $: the trimmed treatment effect estimator is not equal to the treatment effect.
    \item $\Delta$ is monotonic increasing in $\gamma$: the difference between the trimmed treatment effect estimator and the treatment effect increases with the percentage of data trimmed.
\end{itemize}
The goal would be to choose $\gamma$ such as the gap $\Delta$ is minimized while also reducing the estimator's variance.
The issue is that minimizing $\Delta$ and the variance are two conflicting objectives: one is minimized
by trimming less data and the other by trimming more data.

In practice, we should also consider the assumption of the t-test (and Yuen's t-test) that requires the numerator to be
normally distributed.
If the data is highly skewed, we might want to trim more data to make sure the numerator is normally distributed for a
given sample size.

From a business perspective, we want to choose the smallest trimming percentage $\gamma$ because we want to generalize
the results to the largest share of the population possible.
Trimming 40\% of the population on each side essentially makes the inference only valid for the middle 20\% of the
population.

Finally, we want to choose $\gamma$ such that we have enough power to run the experiment for a given sample size.
Increasing $\gamma$ means we loose some generalization but also increases our chance of detecting a significant effect for a
smaller part of the population.

In summary, the choice of $\gamma$ is a trade-off between generalization, power, and making sure that the assumption
of the T-test are met.

\subsubsection{Recommendations}
If you can afford to run simulations prior to each experiment and for each metric of interest, one could use the
results of the simulation to choose the optimal $\gamma$.
The optimal $\gamma$ would be the smallest trimming percentage that reaches the target power for the smallest effect size
that you care about while maintaining the target type I error rate (typically tied to the t-test assumptions).

If you can not afford to run simulations prior to each experiment for each metric, we have seen very good gain with
0.1\% or 1\% trimming.
In our simulations and experience, slightly trimming the population really made a difference on the sensitivity of the
test while not losing much generalization.

\subsubsection{Note about repeated data in the tails}
The proportion of repeated value in a distribution is also something to consider when making a choice for
the trimming percentage.
To make inferences, we have to be careful not to trim too much and end up with only the same value repeated in the whole sample.
The share of repeated values can vary from one experiment to another, making it hard to fully recommend a universally good trimming percentage.

A common example of this problem in practice is with the zero inflated family.
A metric like revenue per website visitor is often dominated by 0s.
We are unable to make an inference if we trim 5\% on each tails of a sample that is 99\% 0s because we are left with only 0s in both groups.

\subsection{Handling Missing Pre-Experiment Data}
There are several reasons why pre-experimental data might be missing for a subject.
This may happen because some subjects are visiting for the first time or because their visits are infrequent, resulting in
their absence during the pre-experiment period.

When missing pre-experiment observations, $\theta$ as defined at ~\eqref{eq:trimmed_cuped_theta} should be computed
using subjects with complete data only.

As the proportion of missing pre-experimental data increases, we want $\overline{Y}^{**}_{tj}$ to converge to
$\overline{Y}_{tj}$.
In other words, the more missing pre-experiment data we have, the smaller the CUPED adjustment should be.
We also need to keep the paired data structure intact prior to trimming which is necessary to compute
the covariance between the trimmed metric and the covariate.
One way to achieve both goals is to impute the pre-experimental data of a subject $X_i$ using the pooled sample trimmed mean
$\overline{X}_t$.
Doing this is the same as not adjusting a subject with missing pre-experiment data.
It is easy to see that if all pre-experiment data is missing for a subject, then $\overline{Y}^{**}_{tj} = \overline{Y}_{tj}$.

%% file: empirical_results.tex
\section{Empirical Results} \label{sec:empirical_results}
The results in this section were obtained by Monte Carlo simulations.
We have three scenarios:
\begin{itemize}
    \item \textbf{Normal}: The Normal scenario is included here as a benchmark.
    This distribution is symmetric and has a short tail.
    These are favorable conditions that are rarely met in practice.
    For the control group in these simulations, we generated data from a normal distribution with a mean of 5 and a standard deviation of 100.
    \item \textbf{Lognormal}: For a more realistic scenario, we consider the lognormal distribution.
    The feature of interest for this scenario is its right-skewness.
    Unlike a symmetrical distribution like the Normal distribution, the trimmed mean is not equal to the mean in this scenario.
    The control group's data in these simulation is randomly sampled from a lognormal distribution parametrized to have a mean of 5 and a standard deviation of 1,000.
    This parametrization allows us to control the skew directly through the standard deviation parameter.
    \item \textbf{Zero inflated Lognormal}: The zero inflated lognormal distribution is an even more challenging case when making inference on the mean.
    Of all three, it most resembles real world metrics like subject expenditure for online e-commerce.
    These metrics involve zero-inflation and strictly positive right-skewed data as most subjects never spend, but a handful spend exceptionally large amounts.
    For the control group in these simulations, we parametrized the distribution to have 90\% zeroes, an overall mean of 5 and standard deviation of 1,000 (zeroes included).
\end{itemize}
Each scenario was simulated 50,000 times.
For each simulation, we generate two groups (control and treatment) of 100,000 subjects each, a sample size found in many online experiments.

The control group's observations are generated from the distribution described in the first column of the result tables.
The treatment group's data are from a same distribution, but with a treatment effect added to the mean.

The pre-experimental data are drawn from the same distribution for both the control and treatment groups.
We use the control group's distribution as the pre-experiment distribution.
For example if the scenario is lognormal, the pre-experiment data for both groups is drawn from the same lognormal distribution as the control group's observations.
This follows the experimental assumption that, prior to exposure to the experiment, subjects are from the same population.
This must also be true for the CUPED adjustment to be unbiased.
We generated pre-experiment values that are correlated with the experimental data using a Gaussian copula ~\cite{copula_modeling} for two spearman correlation $\rho_s$ scenarios.
The spearman correlation corresponds to the covariance specified in the covariance matrix of the copula.

For the Normal scenario the spearman correlation directly maps to the pearson correlation.
For the lognormal and zero inflated lognormal simulations, a spearman correlation $\rho_s$ of 0.95 resulted in a pearson
correlation generally in the range of 0.62 - 0.83, which is a typical range for online experiments under good conditions.
CUPED reduces variance the most in cases like this, where the correlation between the experimental data and the covariate is high.
A spearman correlation $\rho_s$ of 0.25 translates to a pearson correlation of under 0.05 in our simulations.
A Pearson correlation near zero often indicates an ineffective covariate choice or a lack of suitable historical data, such as in experiments with new subjects.
CUPED has little influence in cases like these.

In all simulations, we set $\alpha$ = 0.05.
For Yuen's t-test results, we trimmed the top and bottom 1\% of observations.
We chose this fraction to demonstrate that even with a small trimmed proportion, the benefits are obvious.

\subsection{Methods comparison: Power}
To measure power, the treatment group is generated from the same distribution as the control group, but with its mean shifted by +0.25 which represents a relative treatment effect of +5\%.
Recall that the trimmed effect is not equal to the untrimmed effect when a distribution is not symmetrical.
\autoref{tab:treatent_effect} shows the difference between the true population untrimmed and trimmed treatment effects for the scenarios we simulated.
Smaller effects are usually harder to detect, but as we will see, the reduction in variance more than makes up for the reduction in treatment effect.

\begin{table}[ht]
    \centering
    \caption{Untrimmed and trimmed treatment effects}
    \label{tab:treatent_effect}
    \begin{tabular}{l|rr}
        \toprule
        Scenario & Untrimmed & 1\% trimmed \\
        \midrule
        Normal & .250 & .250 \\
        Lognormal & .250 & .066 \\
        Zero inflated lognormal & .250 & .024 \\
        \bottomrule
    \end{tabular}
\end{table}

\begin{table}[ht]
    \centering
    \caption{Power comparison (\%)}
    \label{tab:power}
    \begin{tabular}{ll|rr|rr}
        \toprule
                &         & \multicolumn{2}{c|}{$\rho_s$ .25} & \multicolumn{2}{c}{$\rho_s$ .95} \\
                &         & Without & With  & Without & With \\
        Scenario & T-test & CUPED   & CUPED & CUPED   & CUPED\\
        \midrule
        \multirow{2}{*}{Normal} & Welch & 8.50 & 8.70 & 8.48 & 43.11 \\
                          & Yuen  & 8.43 & 8.70 & 8.46 & 42.67 \\
        \cline{1-6}
        \multirow{2}{*}{Lognormal} & Welch & 4.46 & 4.41 & 4.35 & 6.47 \\
                             & Yuen  & 64.66 & 64.94 & 64.92 & \textbf{99.34} \\
        \cline{1-6}
        \multirow{2}{*}{Zero inflated} & Welch & 3.89 & 3.80 & 3.82 & 5.24 \\
                  lognormal      & Yuen  & 28.58 & 28.62 & 28.66 & \textbf{66.39} \\
        \cline{1-6}
        \bottomrule
    \end{tabular}
\end{table}

\autoref{tab:power} shows the power obtained with Welch's t-test and Yuen's t-test with or without CUPED.
As expected, Yuen's t-test outperforms in every scenario except the Normal case.
It is established that, for the Normal distribution, the sample mean and the sample standard deviation are the best measures of scale and location ~\cite{McLaughlinTukey1961}.
Thus, the power of Yuen's t-test never exceeds the corresponding power of Student's t under exact normality ~\cite{initial_yuen_test_article_1974} and we notice the reduction of power is very small.
Therefore, from the point of view power, the downside of trimming is minimal even if the data are normally distributed.

In scenarios more common in practice (lognormal and zero inflated lognormal), the power of the Yuen's t-test is largely above Welch's.
The highest power is reached when CUPED is used in conjunction with the Yuen's t-test.
CUPED is only beneficial when the pre-experimental data is highly correlated with the experimental data.
Cases with a spearman correlation $\rho_s$ of 0.25 see little improvement from CUPED, but the power gains are clear when $\rho_s$ = 0.95.

The improvement in power shown in \autoref{tab:power} despite a smaller effect is due to the large reduction in variance.
In \autoref{tab:t_stat_denominator} we present the average denominator of the t-statistic to quantify the uncertainty in the typical statistical test.
Except in the Normal scenario, the CUPED adjusted Yuen's t-test has the lowest denominator, leading to more power as we
saw in \autoref{tab:power}.

\begin{table}[ht]
    \centering
    \caption{Mean T-statistics denominator}
    \label{tab:t_stat_denominator}
    \begin{tabular}{ll|rr|rr}
        \toprule
                 &         & \multicolumn{2}{c|}{$\rho_s$ .25} & \multicolumn{2}{c}{$\rho_s$ .95} \\
                 &        & Without & With  & Without & With \\
        Scenario & T-test & CUPED   & CUPED & CUPED   & CUPED\\
        \midrule
        \multirow{2}{*}{Normal} & Welch & 0.447 & 0.433 & 0.447 & 0.140 \\
                          & Yuen & 0.448 & 0.434 & 0.448 & 0.141 \\
        \cline{1-6}
        \multirow{2}{*}{Lognormal} & Welch & 1.496 & 1.482 & 1.413 & 0.741 \\
                             & Yuen & 0.028 & 0.028 & 0.028 & 0.015 \\
        \cline{1-6}
        \multirow{2}{*}{Zero inflated} & Welch & 1.835 & 1.829 & 1.871 & 0.962 \\
                  lognormal      & Yuen & 0.017 & 0.017 & 0.017 & 0.010 \\
        \cline{1-6}
        \bottomrule
    \end{tabular}
\end{table}

\subsection{Methods comparison: False positive rate}
\autoref{tab:false_positive_rate} shows the false positive rate from Welch's t-test and Yuen's t-test with or without CUPED.
Here, the treatment group's observations are generated from the same distribution as the control group.
This means that both the treatment effect and trimmed treatment effect are 0.

\begin{table}[ht]
    \centering
    \caption{False positive rate (\%)}
    \label{tab:false_positive_rate}
    \begin{tabular}{ll|rr|rr}
        \toprule
                 &         & \multicolumn{2}{c|}{$\rho_s$ .25} & \multicolumn{2}{c}{$\rho_s$ .95} \\
                 &        & Without & With  & Without & With \\
        Scenario & T-test & CUPED   & CUPED & CUPED   & CUPED\\
        \midrule
        \multirow{2}{*}{Normal} & Welch & 4.90 & 4.84 & 4.90 & 4.84 \\
                          & Yuen & 4.82 & 4.75 & 4.87 & 4.84 \\
        \cline{1-6}
        \multirow{2}{*}{Lognormal} & Welch & 3.13 & 3.07 & 3.23 & 3.90 \\
                             & Yuen & 4.82 & 4.93 & 4.83 & 5.00 \\
        \cline{1-6}

        \multirow{2}{*}{Zero inflated} & Welch & 3.05 & 3.12 & 3.28 & 3.94 \\
                       lognormal & Yuen & 4.97 & 4.86 & 4.90 & 4.88 \\
        \cline{1-6}
        \bottomrule
    \end{tabular}
\end{table}

As expected in the lognormal and zero inflated lognormal cases, the false positive rate of Welch's t-test doesn't reach the nominal level.
Welch's t-test becomes more conservative as the tail of the underlying distribution gets longer ~\cite{initial_yuen_test_article_1974}.

On the other hand, Yuen's t-test false positive rate is closer to the nominal $\alpha$ = 0.05
This attractive result regarding the false positive rate is maintained with and without CUPED.

%% file: conclusion.tex
\section{Conclusion}

This paper addressed the difficulties of estimating treatment effects in A/B testing with zero-inflated
and skewed metrics.
Traditional methods like Welch's t-test often fail in these situations because they rely on averages
and assume normality.
We introduced a new approach that blends CUPED's variance reduction with Yuen's t-test, which uses
trimmed means, to improve sensitivity.

Through simulations, we demonstrated that this combined approach achieves higher statistical power and more closely controls the
false positive rates compared to traditional tests.
The benefits are even more pronounced when there is a high correlation between pre-experiment and in-experiment data,
highlighting the effectiveness of CUPED's variance reduction capabilities.

Overall, our contribution facilitates faster and more accurate detection of treatment effects, enabling quicker
decision-making in controlled experiments without compromising statistical power.
This approach expands the practical utility of A/B testing in environments dominated by high-skew and/or high-variance
metrics, providing a foundation for more adaptive methodologies in data-driven decision-making contexts.